\documentclass[12pt,preprint]{aastex}

\usepackage{graphicx}
\usepackage{amsmath}

\def\MNRAS{{\it Mon.\ Not.\ Royal Astron.\ Soc. }}

\def\PASJ{{\it Publ.\ Astron.\ Soc.\ Japan }}
\def\ApJ{{\it Astroph.\ J. }}
\def\ApJL{{\it Astroph.\ J.\ Lett. }}

\def\AA{{\it Astron.\ Astroph. }}

\def\CQG{{\it Clas.\ Quant.\ Grav. }}

\shorttitle{ISS-Lobster--PTA synergy}
\shortauthors{Schnittman}

\begin{document}

\title{Coordinated Observations with Pulsar Timing Arrays and
  ISS-Lobster} 

\author{Jeremy D.\ Schnittman}
\affil{NASA Goddard Space Flight Center \\
Greenbelt, MD 20771}
\email{jeremy.schnittman@nasa.gov}

\begin{abstract}
Supermassive black hole binaries are the strongest gravitational
wave sources in the universe. The systems most likely to be observed
with pulsar timing arrays (PTAs) will have particularly high masses
($\gtrsim 10^9 M_\odot$), long periods ($T_{\rm orb} \gtrsim 1$ yr), and be in
the local universe ($z \lesssim 1$). These features are also the 
most favorable for bright electromagnetic counterparts, which should
be easily observable with existing ground- and space-based
telescopes. Wide-field X-ray observatories such as ISS-Lobster will
provide independent candidates that can be used to lower the threshold
for PTA detections of resolvable binary sources. The primary challenge
lies in correctly identifying and 
characterizing binary sources with long orbital periods, as opposed to
``normal'' active galactic nuclei (AGN) hosting single black holes. Here too
ISS-Lobster will provide valuable new understanding into the wide
range of behaviors seen in AGN by vastly expanding our sample of X-ray
light curves from accreting supermassive black holes. 
\end{abstract}

\keywords{black hole physics -- accretion disks -- X-rays:binaries}

\section{EXECUTIVE SUMMARY}\label{section:summary}

We present here an overview of coordinated electromagnetic (EM) and
gravitational-wave (GW) observations of supermassive black holes
binaries (SMBHBs) with the proposed ISS-Lobster all-sky soft X-ray
imager and the Pulsar Timing
Array (PTA) gravitational wave observatory. The feasibility of
such a program depends on two major unknowns: the rate of SMBHB
mergers in the local Universe, and the nature of associated EM
signatures to such events.

To estimate the merger rates, we combine two relatively
well-constrained observations: the galaxy merger rate at $z\lesssim 1$
and the SMBH mass function over the same time span. The former comes
from a variety of techniques, all of which seem to agree within a
factor of order unity. We adopt the rates given by \citet{lotz:11},
which come from counting merger fractions in deep HST surveys,
combined with theoretical/computational estimates for the amount of
time a system will actually appear to be merging morphologically. The
SMBH mass distribution function comes from X-ray surveys of AGN,
combined with independent estimates of their typical Eddington-scaled
accretion rates and duty cycles \citep{ueda:03,merloni:04}. 

Combined, we find merger rates of $R \sim 3\times 10^{-4}$ Mpc$^{-3}$
Gyr$^{-1}$ for $M\approx 10^8 M_\odot$ and $R \sim 3\times 10^{-5}$
Mpc$^{-3}$ Gyr$^{-1}$ for $M\approx 10^9 M_\odot$. For close binary
systems, where the orbital evolution is governed by gravitational
radiation, the amount of time spent at a given separation $a$ is
proportional to $a^4 \propto T_{\rm orb}^{8/3}$. For SMBHBs with
$T_{\rm orb} \sim 1$ year (typical for PTA sources), the
time-to-merger is still on the order of Myrs. This ensures that, while
the merger {\it rates} are very small, the merger {\it fraction} can be much
larger.  

Out to redshift $z=1$, we expect $N\approx 2\times 10^5$ PTA sources
with $M\sim 10^8 M_\odot$ and $N\approx 400$ for $M\sim 10^9
M_\odot$, which combine to give a characteristic GW strain amplitude
of $h_c \sim 10^{-15}$. A simple rule of thumb for PTA strain sensitivity is
$h\approx \delta t/T_{\rm obs}$, where $\delta t$ is the accuracy of
the pulsar timing residual measurement, and $T_{\rm obs}$ is the
length of the observation. So to detect the expected stochastic signal
at $f\sim$ (1 yr)$^{-1}$, a timing accuracy of 
$\sim 30$ns would be required. 
From the population of thousands of unresolved SMBHBs, there is a
small chance ($\sim \mbox{few }\%$) that one system will be particularly
massive or nearby so that its GW signal rises above the background and
can be individually resolved with PTA alone. 

To estimate the EM signature of a merging SMBHB, we focus on
theoretical models of circumbinary accretion disks. One generic
feature of such a disk is the formation of hot spots near the inner
edge of the disk, regions of large over-densities and
emissivity. This disk flux can get upscattered into the soft X-ray
band via inverse-Compton processes in the surrounding hot corona. Another common feature is the near-ballistic streams of
gas flowing from the inner edge of the circumbinary disk, across a
low-density gap, and colliding with the individual accretion disks
found around each black hole. The X-ray emission from both these
processes is expected to be modulated in time at roughly the orbital
period, with a total flux proportional to the Eddington luminosity.
For PTA-resolvable systems at distances of $\sim 100$ Mpc, the X-ray
flux from the disk+corona alone would almost certainly be observable
by ISS-Lobster with high signal-to-noise. 

Yet even at greater distances where the GW signal alone
might not be resolvable, ISS-Lobster should still be able to detect
the modulated X-ray signal. If we could measure its period and sky
location with high precision, this information could be used to aid in
the PTA data analysis, potentially turning an unresolved source into a
resolved source. If the existence of an EM counterpart can separate
out the stochastic background, PTA timing noise of $\sim 30$ ns should
be good enough to detect massive binaries out to $\sim 350$ Mpc, 
{\it giving a significant chance ($\approx 40\%$) of identifying a resolved
PTA source with total mass $M\approx 10^9 M_\odot$.} Due to its
relative proximity, high luminosity, and long lifetime, such a SMBHB
candidate would be an excellent target for a wide variety of
multi-wavelength follow-up observations.

Out to distances of $D_L\approx 1$ Gpc, ISS-Lobster might identify
as many as 10 massive binaries with periodic light curves, but
essentially no chance to resolve their GW signal with the current PTA
sensitivity. By comparing ISS-Lobster source number counts with
the slope and amplitude of the stochastic GW signal, we may be able
to infer something about the physical mechanisms (i.e., massive gas
disks, three-body scattering, gravitation radiation, etc.) driving SMBHB
evolution in the PTA regime. 

Lastly, even in the absence of unambiguous detections of SMBHBs,
ISS-Lobster will provide long-duration X-ray light curves of roughly 
400 AGN sampled with weekly cadence. Such a data set will be essential
for understanding the broad range of normal AGN activity, and thus
provide a control group against which to compare binary
candidates.

\section{PULSAR TIMING ARRAYS}\label{section:PTA}

Much like EM astronomy, gravitational wave astronomy has historically
been divided into two complementary approaches, ground-based (e.g.,
LIGO/Virgo) and space-based (LISA), roughly corresponding to
stellar-mass and super-massive black hole sources, respectively. In
recent years, a hybrid approach has emerged, that of pulsar timing
arrays. PTAs used ground-based radio telescopes to measure the effect
of GWs on the propagation of radio pulses from millisecond pulsars
distributed throughout the galaxy. The GWs impart a characteristic
delay on the arrival times of the pulses at Earth. By comparing the
arrival times with an underlying model for the
pulsar/Earth/Sun system, a ``residual'' is measured. The timing
residuals from different pulsars are
correlated in a very well-defined way, providing additional angular
information about any potential underlying GW signal
[\citet{hellings:83}, see \citet{hobbs:10} for a review of the current
instruments and future sensitivity projections]. 

The predominant sources for PTAs are supermassive black hole binaries
at cosmological
distances. These sources can generally be divided into two categories:
stochastic and resolved. This is primarily a data analysis, rather
than astrophysical, distinction. By most estimates, the stochastic
population is more likely to be detected by first-generation PTAs. 
The resolved population---if detected---would simply be the
brightest subset of the total population of SMBHBs throughout the
universe. In practice, they are more massive, relatively
closer, and at higher GW frequencies (and thus closer to merger) than the
average SMBHB. Table 1 shows the typical order-of-magnitude properties
of systems in these two source classes. 

\begin{table}[ht]
\caption{\label{table1} Typical properties of SMBHBs making up
  the stochastic and resolved populations of PTA sources.}
\begin{center}
\begin{tabular}{lcc}
 & stochastic & resolved \\
\hline
$M_{\rm BH}$ ($M_\odot$) & $10^8$ & $10^9$ \\
$f_{\rm GW}$ (Hz) & $10^{-8}$ & $10^{-7}$ \\
$q = M_2/M_1$ & 0.1-1 & 1 \\
$D_L$ (Gpc) & 10 & 0.1 \\
GW strain $h_c$ & $10^{-19}$ & $10^{-14}$ \\
\end{tabular}
\end{center}
\end{table}

Since we expect any gas-powered electromagnetic (EM) counterparts
to be limited to roughly Eddington luminosities, both these
features (high mass and small distance) work in our favor to make
the resolved population of PTA sources bright enough to see easily
with EM telescopes. Furthermore, working under the assumption that
the dynamics of galaxy mergers promote more gas flow into the central
regions, we expect SMBHBs to have higher average accretion rates
relative to isolated AGN \citep{treister:12}. While the most massive
BHs will likely reside in massive, gas-poor elliptical galaxies, this
clearly does not preclude the possibility of quasar activity
\citep{kirhakos:99}. 

\section{EM MECHANISMS}\label{section:em}

An exhaustive review of potential electromagnetic signatures of SMBH
mergers is provided in \citet{schnittman:13} [for recent work on the
  subject, and particularly for PTA-relevant sources, see
  \citet{tanaka:12,sesana:12,tanaka:13,burke:13,dorazio:13,farris:14}]. Figure
1 (reproduced from Schnittman 2013) shows a sample of the mechanisms
that could produce an EM signature. They can be roughly
divided into direct and indirect evidence for mergers, or
alternatively, ``counterparts'' and ``signatures.'' In this discussion
we will use the term counterpart to refer to an EM signal that is
detected (or at least detectable) along with a concurrent GW signal. A
signature is either a direct or indirect EM signal that has no
associated GW detection (we do not consider the GW stochastic
background to be ``associated'' with any specific EM signal).

\begin{figure}[ht]
\begin{center}
\includegraphics[width=0.8\textwidth]{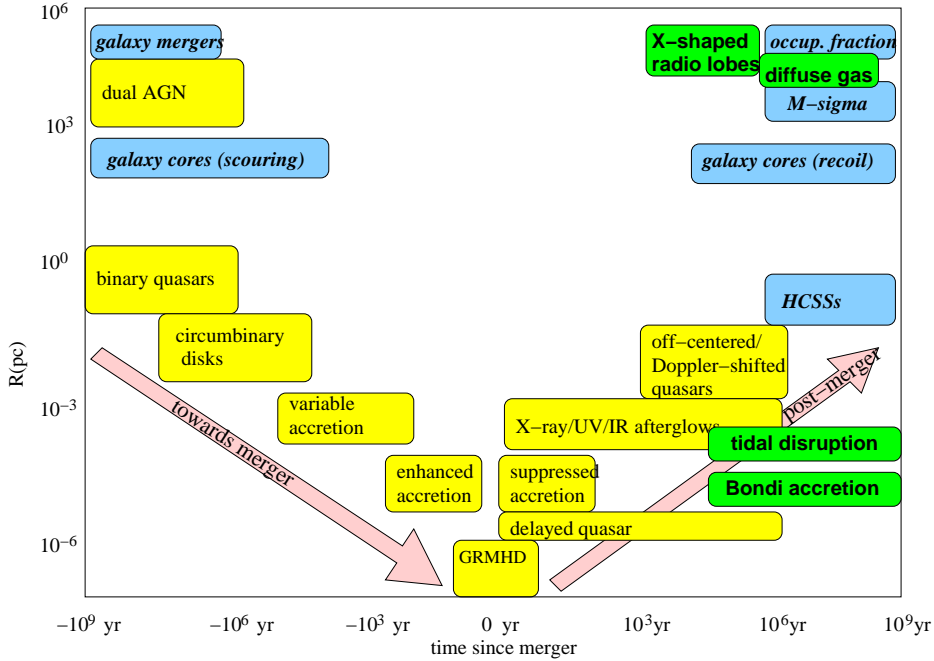}
\caption{\label{source_chart} Selection of potential EM sources,
  sorted by timescale, typical size of emission region, and physical
  mechanism (blue/{\it italic} = stellar; yellow/Times-Roman =
  accretion disk; green/{\bf bold} = diffuse gas/miscellaneous). The
  evolution of the merger proceeds from the upper-left through the
  lower-center, to the upper-right.}
\end{center}
\end{figure}

The axes in Figure 1 are distance from merger vs time before/after
merger. The sources evolve from the upper-left through the
lower-center, to the upper-right of the figure. For the most part, EM
counterparts come from the lower-left/lower-center regions where GW
emission is strongest and most likely to be detected with PTA
observations. These sources all rely on the presence of relatively
dense, cool gas in the form of accretion flows around the two black
holes. 

Thus the study of EM counterparts to SMBHB GW sources is fundamentally
one of understanding the behavior of circumbinary accretion
disks. More specifically, we would like to understand how the behavior
of circumbinary disks differs from that of active galactic
nuclei (AGN) driven by single black holes. We have decades of
observations of AGN across many wavelengths, with ever-improving
spatial, spectral, and temporal resolution. As evidenced by the
multi-dimensional zoological classification schemes (e.g., Antonucci
2011) needed to describe the variety of AGN observations, it is
notoriously hard to say exactly what a ``normal'' AGN looks like. 

Most 1D analytic models of circumbinary disks suggest that the
additional outward torques that the binary imposes on the disk will
arrest significant accretion, and thus EM luminosity, leaving an
evacuated region around the two black holes \citep{pringle:91,
  milos:05}. However, virtually all 2D and 3D simulations of
circumbinary disks show significant accretion across the inner gap,
generally along two gas streams as seen if Figure 2. Other generic
features include a low-density region with radius $R_{\rm in}\sim 2a$,
with $a$ the semi-major axis of the binary, eccentric orbits for the
disk, and spiral density waves
excited in the inner disk. These density waves in turn impart a
gravitational torque on the binary and cause it to shrink [however,
recent work suggests that {\it retrograde} circumbinary disks may show
very different behavior and evolution
\citep{nixon:11,roedig:14a,bankert:14,schnittman:14}].

\begin{figure}[ht]
\begin{center}
\includegraphics[width=0.95\textwidth]{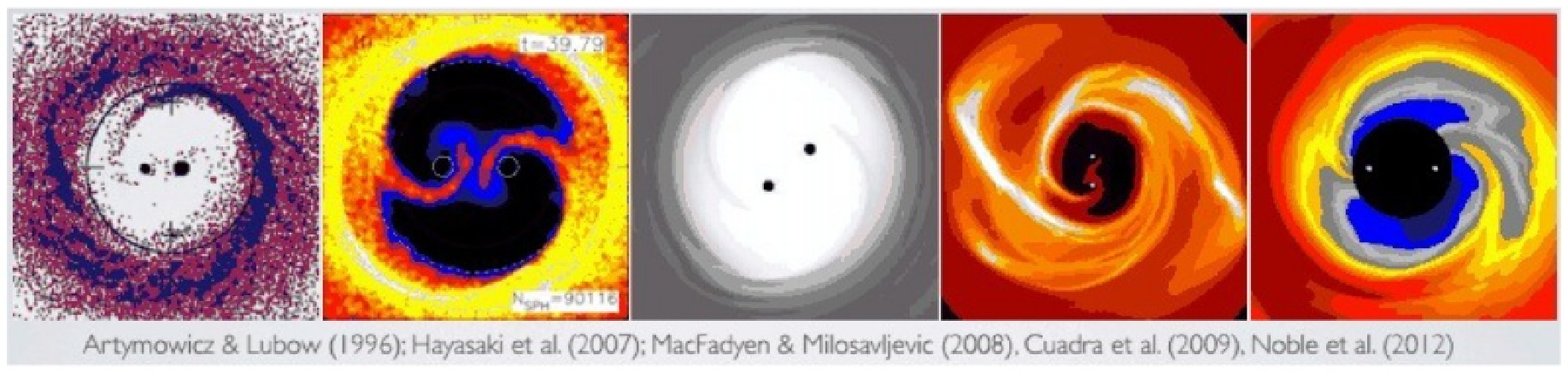}
\caption{\label{disk_pictures} Snapshots from a variety of 2D and 3D
  simulations of circumbinary disks, taken from
  \citet{artymowicz:96,hayasaki:07,macfadyen:08,cuadra:09,noble:12}.}
\end{center}
\end{figure}

In addition to these features, recent grid-based 3D
magneto-hydrodynamic (MHD) simulations by \citet{shi:12,noble:12} show
the presence of a strong $m=1$ density perturbation at radius
$R\approx 2.5a$, shown in Figure
3. The beat frequency between the lump's orbital period and that of
the binary leads to a strong modulation in the disk luminosity at
frequency $\Omega\approx 1.5\Omega_{\rm bin}$, corresponding to each
of the binary members gravitationally exciting the region of
over-density, in turn leading to a raised level of local emission. In
addition to the variability in the total luminosity, there should also
be a variation in the measured flux as seen by an inclined observer,
due to the Doppler shifting of the hot spot moving towards and away
from the observer, with frequency corresponding to the Keplerian
frequency of the orbiting lump $\Omega_{\rm lump} \approx 0.25
\Omega_{\rm bin}$ \citep{noble:12}. 

For the typical masses and densities for AGN disks, the disk should be
optically thick and emitting primarily in the optical/UV.
In almost every AGN observed in the X-ray, there appears to be a hot,
tenuous corona surrounding the accretion disk, leading to
inverse-Compton scattering of the thermal seed photons up to $\sim 100$
keV with luminosity of roughly $10\%$ of the total bolometric
luminosity. So even if the lump is fundamentally a disk feature, it
should still show up in the X-ray light curves due to the local
inverse-Compton emission immediately above the lump. \citet{noble:12}
find that the global luminosity variability has an amplitude of $\sim
5-10\%$. For an observer at inclination $i$, the
Doppler-shifted intensity will scale roughly like $(\delta\rho/\rho)
(1+\sin i\, v/c)^3$. For a circular orbit with binary separation of
$a$ and canonical observer with $i=60^\circ$, this
corresponds to $\delta I/I\sim 2a^{-1/2}$. 

\begin{figure}[ht]
\begin{center}
\includegraphics[width=0.5\textwidth]{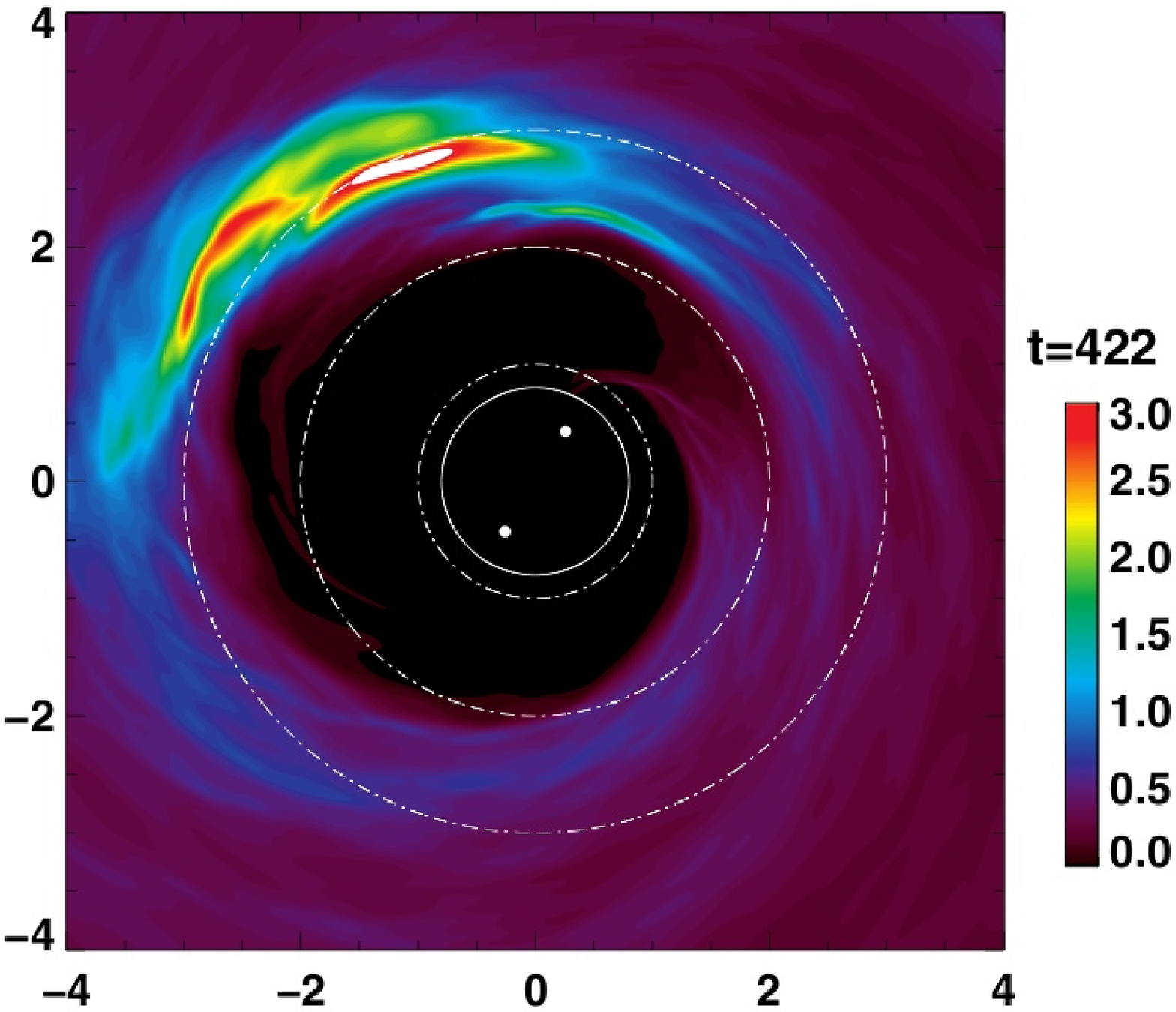}
\includegraphics[width=0.4\textwidth]{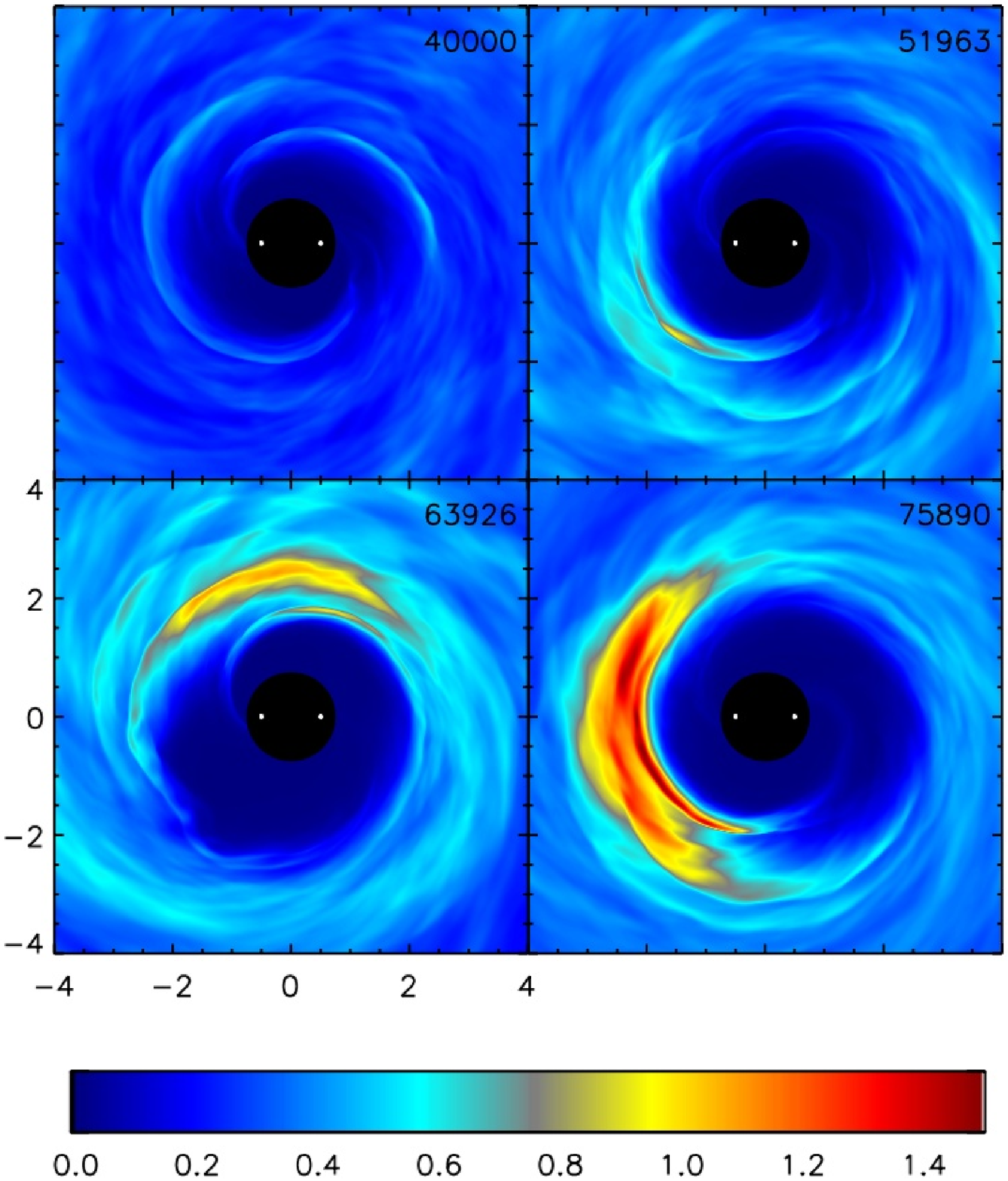}
\caption{\label{shi_noble} Surface density from two different MHD
  simulations of circumbinary disks. {\it left}: Shi et al.\ (2012);
  {\it right}: Noble et al.\ (2012). In both cases, over-dense
  ``lumps'' are formed near the inner edge of the disk.}
\end{center}
\end{figure}

What binary separations might we expect for the PTA sources? From
Kepler's first law, we can write
\begin{equation}\label{a_rg}
a = 100 r_g \left(\frac{T_{\rm orb}}{\mbox{1 yr}}\right)^{2/3}
\left(\frac{M}{10^9 M_\odot}\right)^{-2/3}\, ,
\end{equation}
where $M$ is the total mass of the binary and $r_g=GM/c^2$ is the
gravitational radius, the standard unit of length in the system. From
this expression, we see that when the binary orbital period is one
year, the system is already entering the relativistic regime with $v/c
\sim 0.1$, or an X-ray modulation of $10-20\%$. 

\citet{roedig:14} recently explored two specific spectral
signatures of a circumbinary disk: a low-energy notch in the spectrum
and a high-energy excess, shown in Figure 4. Both signatures
assume the existence of individual accretion disks around each black hole,
something that is typically very difficult to resolve numerically in
simulations. Assuming these ``mini-disks'' do exist, they will likely
extend out to some fraction of the semimajor axis $a$, inside of which
they should behave like normal disks, with gravitational binding
energy being released through gas accretion via the magneto-rotational
instability. Similarly, well outside of the gap region the circumbinary
disk should behave like a regular accretion disk. Since the
temperature of a classical disk scales like $R^{-3/4}$, each radius
contributes to a different part of the continuum spectrum. By removing
the gas from the gap region between the inner mini-disks and the outer
circumbinary disk, a corresponding notch will be removed from the
spectrum. This is shown in Figure 4, reproduced from
\citet{roedig:14}, where $T_0$ is the temperature of the inner edge of
the circumbinary disk. The exact location and shape of this notch
feature will depend on the binary separation, accretion rate,
black hole masses, and mass ratio. 

\begin{figure}[ht]
\begin{center}
\includegraphics[width=0.35\textwidth,angle=90]{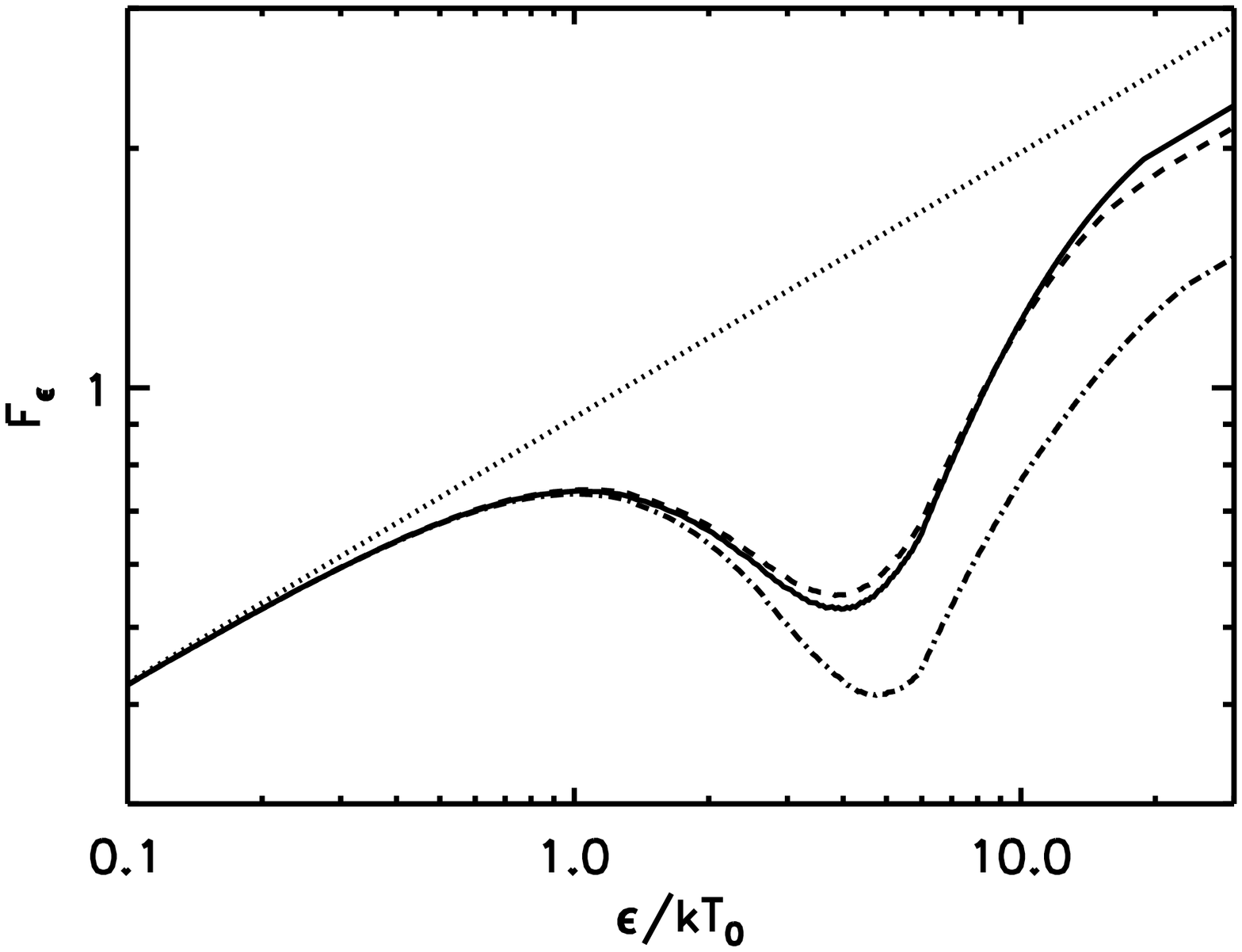}
\includegraphics[width=0.35\textwidth,angle=90]{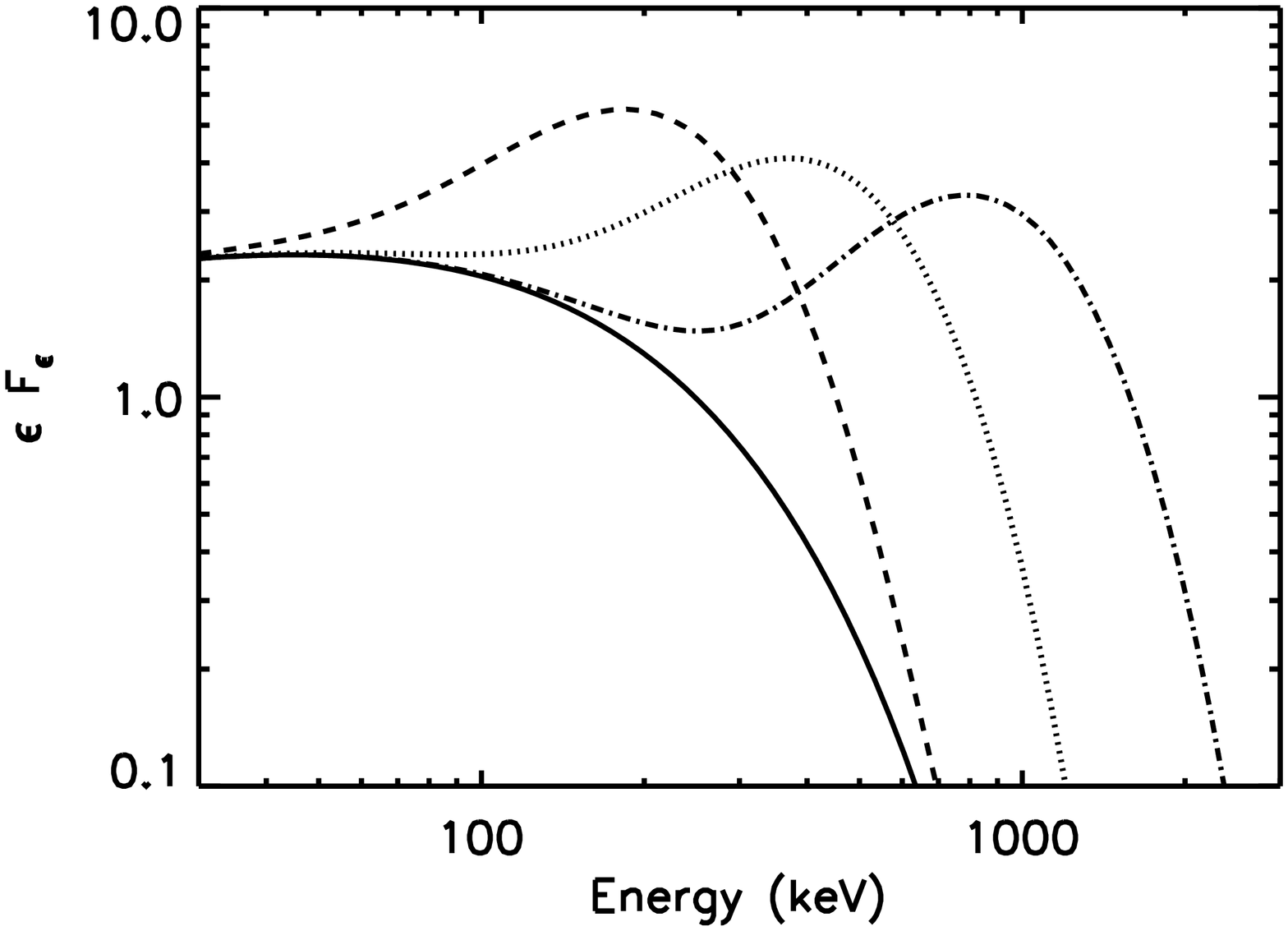}
\caption{\label{roedig14} Spectra of circumbinary disk AGN, with the
  left figure showing the missing continuum emission from the gap
  (scaled to the thermal temperature $T_0$ at the inner edge of the
  disk), and the right figure showing the high-energy emission caused
  by streams shocking on the inner accretion ``mini-disk.'' The
  different curves in each plot represent a range of reasonable model
  parameters, along with the coronal emission expected from a normal
  AGN ({\it right, solid curve}). [reproduced from \citet{roedig:14}]}
\end{center}
\end{figure}

Also shown in the right panel of Figure 4 is the high-energy emission
expected from the 
gas streaming from the outer disk into the inner disks and forming
high-temperature shocks. With stream velocities comparable to the
binary orbital velocity, post-shock temperatures will be of the order $kT
\sim GM/a$, with fiducial values of $\sim 200$ keV for separation of
$a\sim 10^3 r_g$. This hot gas will generally cool rapidly through
pair production and inverse-Compton scattering of seed photons from
the mini-disks, producing a modulated hard X-ray
signal with luminosity proportional to mass flow across the gap. For
eccentric disks (or binaries), modulations of the accretion rate on
the order of $10\%$ or higher should be common
\citep{macfadyen:08,shi:12,noble:12}.

In the shock-stream model, the mini-disks around each black hole
should also power a corona, and thus contribute to the X-ray
flux. Since the majority of potential energy is extracted in the
inner-most region of the disk, it is in fact quite likely that the
mini-disks dominate the total X-ray flux from the system. In that
scenario, the hard X-ray flux ($\gtrsim 100$ keV) will come predominately
from the shock points, while the softer X-rays ($\sim 1-10$ keV) will
come from inverse-Compton scattering of the mini-disk photons off of
mini-coronas. 

If the mini-disks do in fact dominate the emission,
then the chance of detection with ISS-Lobster will be
greater, but the variable fraction of the signal may be smaller,
depending on the inflow timescales of the mini-disks. If the inflow
time is long compared to the orbital period, then the mini-disks will
absorb the variable accretion across the gap and damp out any
modulation in the light curve. However, if the inflow time is
comparable to the orbital period, the mini-disk emission should vary
with the same period and amplitude as the accretion streams. In both
cases, there should still be relativistic Doppler beaming of the
mini-disk flux, again leading to variability amplitudes on the $\sim
10-20\%$ level.

It should be noted that all of these emission mechanisms are nearly
isotropic, further enabling joint GW-EM observations. This is in
marked contrast to the leading candidates for LIGO EM counterparts,
the highly relativistic, beamed jets from gamma-ray bursts
(GRBs). While kilonovae counterparts would be largely isotropic, they
would also have significantly lower flux, making them much harder to
detect with EM observations \citep{metzger:12}. Additionally, because
the PTA GW sources are continuous and long-lived, signal-to-noise will
grow steadily in time along with the X-ray light curves. 

\section{RATES}\label{section:rates}

Traditionally, event rates for space-based gravitational wave
detectors like LISA have been based on large-scale N-body cosmological
simulations tracking the mergers of galaxies and their central SMBHs
from the early Universe through today. This is necessary because many
of the sources will likely be at high redshift $z \gtrsim 5-10$ where
there are no solid observational constraints for the event rates
\citep{sesana:05,sesana:07}. While this approach should also work in
principle for PTA sources \citep{sesana:08}, because they tend to be
much closer (see Table 1), we do not need to rely on purely
theoretical calculations of their rates.

EM observations of galaxy merger rates at $z \lesssim 1.5$ should be
sufficient estimates of SMBH merger rates relevant to PTA observations
(and probably more accurate 
than N-body simulations). \citet{lotz:11} find the major merger rate
($1/4<M_2/M_1<1$) this range to be $\sim 0.1$ Gyr$^{-1}$ per galaxy,
and the minor merger rate ($1/10<M_2/M_1<1/4$) to be roughly three
times that. Combined with the black hole mass function derived from
X-ray surveys such as \citet{ueda:03} and \citet{merloni:04}, we get merger rates of 
$R \sim 3\times 10^{-4}$ Mpc$^{-3}$ Gyr$^{-1}$ for $M\approx 10^8
M_\odot$ and $R \sim 3\times 10^{-5}$ Mpc$^{-3}$ Gyr$^{-1}$ for
$M\approx 10^9 M_\odot$. This ``back-of-the-envelope'' approach agrees
reasonably well with a more detailed analysis by \citet{sesana:13a}. 

To estimate the number of sources in the sky at any one time, one
needs to multiply the rate times the time that the source spends in
that phase. For SMBHBs in the GW-dominated phase of evolution, the
characteristic time spent at a binary separation $a$ is
\citep{peters:64} 
\begin{equation}
t_{\rm GW} = \frac{a}{\dot{a}} = \frac{5}{64}\frac{c^5}{G^3 M^3}
\frac{(1+q)^2}{q} a^4\, ,
\end{equation}
where $M$ is the total mass and $q$ is the mass ratio.
Combining with equation (\ref{a_rg}), and for a fiducial mass ratio of
$q=1/3$, the timescale $t_{\rm GW}$ spent 
with GW period around 1 yr is $\sim 5\times 10^6$ yr for $M=10^8
M_\odot$ and $\sim 10^5$ yr for $10^9 M_\odot$. This corresponds to a
total number of PTA sources out to $z=1$ of $N(M>10^8 M_\odot) \approx
2\times 10^5$ and $N(M>10^9 M_\odot) \approx 400$. 

\section{OBSERVABILITY}\label{section:observability}

The characteristic (i.e., averaged over sky location and source
orientation) strain amplitude from a binary source at luminosity
distance $D_L$ and observed frequency $f_{\rm GW}=2 T^{-1}_{\rm
  orb} (1+z)^{-1}$ is given by
\begin{equation}\label{eqn:hc}
h_c = \frac{G\mathcal{M}}{c^2 D_L}
\left(\frac{G\pi f_{\rm GW} \mathcal{M}}{c^3}\right)^{2/3}\, ,
\end{equation}
where the chirp mass is defined as $\mathcal{M}=(M_1
M_2)^{3/5}M^{-1/5}=q^{3/5}(1+q)^{-6/5}M$. At a redshift of $z=1$, this
gives a characteristic strain of
\begin{equation}
h_c(z=1) \approx 5\times 10^{-17} \frac{q}{(1+q)^2}
\left(\frac{M}{10^9 M_\odot}\right)^{5/3}
\left(\frac{T_{\rm orb}}{\mbox{1 yr}}\right)^{-2/3}\, .
\end{equation}

Again using $q\sim 1/3$, and summing over the incoherent populations
of binaries out to $z=1$, we find 
\begin{equation}
h_c(M>10^{8.5} M_\odot) \sim 3\times 10^{-16}
\end{equation}
and
\begin{equation}
h_c(M<10^{8.5} M_\odot) \sim 1.5\times 10^{-16}\, .
\end{equation}
These very rough estimates are actually in quite close agreement with
those predicted by \citet{sesana:13a,sesana:13b}, using Monte Carlo
realizations of the Universe, based on observed galaxy merger rates and
black hole distribution function out to $z\sim 1.5$. Furthermore, this
stochastic signal should be within reach of reasonable PTA sensitivity
estimates within the next 5-10 years \citep{sesana:13a,sesana:13b}. 

For a single source to rise up above the background, let us
estimate that its strain should be roughly twice that of the
stochastic signal [However, careful data analysis
techniques such as matched filtering may allow for the identification
of an individual source with signal strength comparable to the
background, thereby increasing the chance of detection with time.
(A.\ Sesana, private communication)]. For $M=10^9 M_\odot$, this
corresponds to a 
distance of $D_L \lesssim 100$ Mpc, or a co-moving volume
$3\times10^4$ times smaller than that occupied by the stochastic
population. The expected number of binaries in this
volume with orbital periods of $\sim 1$ year is only $\sim 0.01$, so
the chance of detecting a resolved GW source with PTA alone is maybe a
few percent in the optimistic limit. 

We may get lucky with a particularly
massive binary with $M>10^{10} M_\odot$, but the black hole
distribution function at the high end of the mass range is poorly
known, so it is very difficult to make even rough estimates.
Similarly, pushing to higher frequencies and more equal mass ratios
will increase the distance at which point the sources are resolved,
but also move into the regime where PTA is less sensitive (the
observable strain scales like $h\propto f$ at high frequency).

On the other hand, by extending to lower frequencies (i.e., longer
observations), the number of sources could be increased by a factor of
$(a/a_{\mbox{1 yr}})^4 = (T_{\rm orb}/\mbox{1 yr})^{8/3}$. Of course,
more sources will make the stochastic signal stronger, making it more
difficult to resolve individual sources. At the same time, by going to
very long-period orbits, there is a very good chance that the binary
separation will move out of the purely GW-driven evolution and begin
to probe additional larger-scale astrophysical mechanisms such as
disk-driven migration and three-body relaxation with surrounding
stars \citep{begelman:80,kocsis:11,sesana:13b,ravi:14}.

What about electromagnetic observability? Let us first consider the
less uncertain model of emission from the circumbinary disk. In
normal accretion disks, the primary source of EM power is the
conversion of gravitational potential energy into turbulent magnetic
stress within the disk body \citet{balbus:91}. This is in turn converted to both thermal
disk radiation and high-energy X-rays via inverse-Compton scattering
in the corona \citep{noble:11,schnittman:13a}. The radiative
efficiency of such a disk is roughly $1-E_{\rm ISCO}$, where $E_{\rm ISCO}$ is
the specific energy of the gas at the inner edge of the accretion
disk. In the Newtonian limit, $E_{\rm ISCO} = c^2/(2r_{\rm ISCO})$
with $r_{\rm ISCO}$ measured in gravitational radii. For a
circumbinary disk with inner edge $r_{\rm in} = 2a$, the radiative
efficiency is reduced by a factor of $r_{\rm ISCO}/(2a)$. 

If the gas is supplied at large radius at $\dot{m}$ times the Eddington
accretion rate, the EM luminosity of the disk alone will be
\begin{equation}
L_{\rm disk} = \dot{m} L_{\rm Edd} \left(\frac{r_{\rm isco}}{2a}\right)
\approx 2\times 10^{45}\, \dot{m} \left(\frac{M}{10^9 M_\odot}\right)^{5/3}
\left(\frac{T_{\rm orb}}{\mbox{1 yr}}\right)^{-2/3} \mbox{ erg/s}\, .
\end{equation}
Typical AGN have X-ray fractions of the total bolometric luminosity of
$f_x \sim 0.1$ in the ISS-Lobster band. For GW-resolvable sources at
$D_L=100$ Mpc, the X-ray flux will be
\begin{equation}\label{eqn:F_x}
F_x = \dot{m} \frac{f_x}{0.1} 10^{-11}\, \mbox{erg/s/cm}^2\, ,
\end{equation}
with a modulation of $\sim 20\%$ on timescales of the orbital
period. 

For the shocked streams discussed above, $f_x$ would likely be
significantly smaller, since most of the associated emission is
limited to energies above 10 keV. 
Including the emission from the mini-disks/coronae will increase the X-ray
flux by another factor of $\sim (a/r_{\rm ISCO})$, with comparable
levels of variability. This would mean equation (\ref{eqn:F_x}) could
hold for sources as far as $D_L\approx 1$ Gpc! In this volume, there
could be as many as $\sim 10$ SMBHBs in the PTA band with total mass
greater than $10^9 M_\odot$. 

ISS-Lobster is expected to reach flux levels of $10^{-11}$
erg/s/cm$^2$ for the entire sky scan each day. This means that any
source that is potentially resolvable with PTA observations should be
easily observable with ISS-Lobster, as long as it is accreting a
significant amount of gas. Large scale gas flows and accretion have
long been predicted as generic features of galaxy mergers, and there
is mounting observational evidence that there is in fact a strong
correlation between galaxy mergers and AGN activity
\citep{koss:12}. Not only do galaxy mergers drive gas inflow, but gas
inflow in turn drives black hole mergers \citep{cuadra:09}, further
increasing the likelihood that SMBHBs will be formed and observed in
the presence of circumbinary disks with relatively high values of $\dot{m}$.

On a weekly timescale, ISS-Lobster should be able to sample roughly
400 AGN in the sky with fluxes greater than $\approx 2\times 10^{-12}$
erg/s/cm$^2$ \citep{hasinger:05}. From the event rates calculated 
above, we estimate there is a moderate chance that one of these
objects is in fact a SMBHB in the PTA band, but too far away to
resolve from the GW signal alone. However, with the sky location and
precise orbital period provided by ISS-Lobster, it is quite possible
that the GW signal could be picked out from the stochastic
background. If in fact we could separate the individual signal from
the background, equation (\ref{eqn:hc}) suggests that current PTA
detectors should be sensitive out to $D_L \approx 350$ Mpc for $M=10^9
M_\odot$. Within this range, there is $\approx 40\%$ chance of such a
SMBHB existing. The range for $M=10^8 M_\odot$ is only $D_L \approx 6$
Mpc, giving almost no chance of resolving a PTA source (which is why
they are in the stochastic category in the first place!). 

For the canonical stochastic PTA sources, the X-ray flux would be on
the order of $\sim 10^{-15}$ erg/s/cm$^2$. While far too faint for
ISS-Lobster, these objects would
certainly show up in the Chandra deep field surveys, but
identification as periodically variable sources would be nearly
impossible.

\section{DISCUSSION}\label{section:discussion}

We must end this discussion the same way it began, with a clear
disclaimer: the
rates and properties of the SMBHB sources discussed above are
intentionally very rough approximations, accurate to an order of
magnitude at best. We simply do not believe greater precision is
justified or even particularly helpful at this stage in the
field. Yet even these rough theoretical estimates are more than enough to
justify intense and comprehensive observational campaigns, which will
hopefully yield actual detections, or at the very least, more reliable
limits. 

From a variety of independent methods (N-body cosmological
simulations, deep field HST and ground-based surveys, local Universe
AGN surveys), the merger rate of galaxies and their central black
holes appears to be roughly one major---and a few minor---mergers per
galaxy per Hubble time. This implies several hundred (thousand) SMBHBs
with total masses $\sim 10^9 M_\odot$ ($10^8 M_\odot$), orbital
periods of roughly one year, and within the volume out to
$z=1$. Combined, these thousands of sources contribute to a stochastic
background with characteristic strain amplitude of $h_c \lesssim
10^{-15}$. While typical isolated SMBHs have relatively low duty
cycles, it is believed that the galactic merger process
will make it much more likely that the SMBHB sources are accreting gas
at nearly the Eddington rate, leading to particularly bright and
variable X-ray sources.

There is a small chance that a few SMBHB sources might lie
sufficiently close or be sufficiently massive such that their GW
signal rises above the stochastic background and can be individually
resolved with PTA observations. If these systems also have
circumbinary accretion disks, they will almost certainly be detectable
with high signal-to-noise with ISS-Lobster. Such a combined EM-GW
signal would greatly raise the significance of both measurements, and
undoubtedly lead to a massive multi-wavelength targeted observation
campaign. 

There is a larger number of sources that should be within the
detection limits of ISS-Lobster, but only resolvable with PTA if the
sky position and frequency are known {\it a priori}. Thus the
ISS-Lobster--PTA synergy has great potential to significantly enhance 
both science programs. To a certain extent, this is the converse of
the analogous LIGO-GRB connection, where many off-axis GW sources will
be seen by LIGO with high SNR, but go undetected in the EM spectrum.

Even for the unresolvable sources, there is much to be learned from
coordinated ISS-Lobster--PTA science. In particular, the high-cadence
observations of roughly 400 AGN in the X-ray band will expand our
sample of long-baseline AGN light curves by nearly two orders of
magnitude. Time-domain observations of a few of the brightest AGN have shown us that there is
a very wide range of variability behavior from single accreting black
hole systems. Thus it is imperative to fully explore the range of
``normal'' AGN before we have a convincing chance to identify a given
source as a binary from the light curve alone. ISS-Lobster will do
exactly this, and also provide promising targets for deeper, targeted
multi-wavelength observing campaigns of the most interesting variable
sources, much like has been done for BAT-selected AGN studies such as
\citet{koss:12}. 

Similarly, the amplitude and slope of the unresolved stochastic GW
background will provide important limits on the evolution of SMBHB
systems. The balance of gas-driven migration, stellar scattering, and
gravitational radiation losses all combine to give a complex picture
of the evolution of merging black holes and their environments
\citep{ravi:14}. By
including population information from ISS-Lobster (as well as
complementary surveys like the Chandra deep field and XMM-Newton
COSMOS field), we will have
constraints to theoretical models of binary evolution. For example, if
we detect many bright X-ray sources, but a relatively low GW signal,
it might indicate gas-driven evolution well beyond the conventional stage
where the disk decouples from the binary \citep{milos:05}. On the
other hand, a flat GW spectrum, coupled with a lack of periodic X-ray
sources, might suggest that high-eccentricity binaries are common and
the gas disk decouples at an early stage \citep{schnittman:14}.

\noindent
{\bf Acknowledgments:} We would like to thank Jordan Camp, Cole
Miller, and Alberto Sesana for many useful comments. This work was
supported in part by NASA grant ATP12-0139.

\newpage


\begin{thebibliography}{99}
\bibitem[Antonucci (2011)]{antonucci:11} Antonucci, R. 2011,
  arXiv:1101.0837 
\bibitem[Artymowicz \& Lubow(1996)]{artymowicz:96} Artymowicz, P., \&
  Lubow, S.\ H. 1996 \ApJL {\bf 467}, 77
\bibitem[Balbus \& Hawley(1991)]{balbus:91} Balbus, S.\ A., \& Hawley,
  J.\ F. 1991, \ApJ {\bf 376}, 214 
\bibitem[Bankert et al.(2014)]{bankert:14} Bankert, J., Krolik,
  J.\ H., \& Shi, J.-M. 2014, \ApJ submitted
\bibitem[Begelman et al.(1980)]{begelman:80} Begelman, M.\ C.,
  Blandford, R.\ D., \& Rees, M.\ J. 1980 {\it Nature} {\bf 287}, 307
\bibitem[Burke-Spolaor(2013)]{burke:13} Burke-Spolaor, S. 2013, \CQG
  {\bf 30}, 224013
\bibitem[Cuadra et al.(2009)]{cuadra:09} Cuadra J., Armitage P.\ J.,
  Alexander, R.\ D., \& and Begelman, M.\ C. 2009 \MNRAS {\bf 393}
  1423
\bibitem[D'Orazio et al.(2013)]{dorazio:13} D'Orazio, D.\ J., Haiman,
  Z., \& MacFadyen, A. 2013, \MNRAS {\bf 436}, 2997
\bibitem[Farris et al.(2014)]{farris:14} Farris, B.\ D., Duffell, P.,
  MacFadyen, A.\ I., \& Haiman, Z. 2014, \ApJ {\bf 783}, 134
\bibitem[Hasinger et al.(2005)]{hasinger:05} Hasinger, G., Miyaji, T.,
  \& Schmidt, M. 2005, \AA {\bf 441}, 417
\bibitem[Hayasaki et al.(2007)]{hayasaki:07} Hayasaki, K., Mineshige,
  S., \& Sudou, H. 2007, \PASJ {\bf 59}, 427
\bibitem[Hellings \& Downs(1983)]{hellings:83} Hellings, R.\ W., \&
  Downs, G.\ S. 1983, \ApJL {\bf 265}, L39 
\bibitem[Hobbs et al.(2010)]{hobbs:10} Hobbs, G., et al. 2010, \CQG
  {\bf 27}, 084013
\bibitem[Kirhakos et al.(1999)]{kirhakos:99} Kirhakos, S., Bahcall,
  J.\ N., Schneider, D.\ P., \& Kristian, J. 1999, \ApJ {\bf 520}, 67
\bibitem[Kocsis \& Sesana(2011)]{kocsis:11} Kocsis, B., \& Sesana,
  A. 2011, \MNRAS {\bf 411}, 1467
\bibitem[Koss et al.(2012)]{koss:12} Koss, M., et al. 2012, \ApJL {\bf
  746}, 22
\bibitem[Lotz et al.(2011)]{lotz:11} Lotz, J.\ M., Jonsson, P., Cox,
  T.\ J., Croton, D., Primack, J.\ R., Somerville, R.\ S., \& Stewart,
  K. 2011, \ApJ {\bf 742}, 103
\bibitem[MacFadyen \& Milosavljevic(2008)]{macfadyen:08} MacFadyen,
  A.\ I., \& Milosavljevic, M. 2008, \ApJ {\bf 672} 83
\bibitem[Merloni (2004)]{merloni:04} Merloni, A. 2004, \MNRAS {\bf
  353}, 1035
\bibitem[Metzger \& Berger(2012)]{metzger:12} Metzger, B.\ D., \&
  Berger, E. 2012, \ApJ {\bf 746}, 48
\bibitem[Milosavljevic \& Phinney(2005)]{milos:05} Milosavljevic, M.,
  \& Phinney, E.\ S. 2005 \ApJL {\bf 622}, 93
\bibitem[Nixon et al.(2011)]{nixon:11} Nixon, C.\ J., Cossings,
  P.\ J., King, A.\ R., \& Pringle, J.\ E. 2011, \MNRAS {\bf 412},
  1591
\bibitem[Noble et al.(2011)]{noble:11} Noble, S.\ C., Krolik, J.\ H.,
  Schnittman, J.\ S., \& Hawley, J.\ F. 2011, \ApJ {\bf 743}, 115
\bibitem[Noble et al.(2012)]{noble:12} Noble, S.\ C., Mundim, B.\ C.,
  Nakano, H., Krolik, J.\ H., Campanelli, M., Zlochower, Y., \& Yunes,
  N. 2012, \ApJ {\bf 755}, 51
\bibitem[Peters(1964)]{peters:64} Peters, P.\ C. 1964, Phys.\ Rev.\
  {\bf 136}, 1224
\bibitem[Pringle(1991)]{pringle:91} Pringle, J.\ E. 1991, \MNRAS {\bf
  248}, 754
\bibitem[Ravi et al.(2014)]{ravi:14} Ravi, V., Wyithe, J.\ S.\ B.,
  Shannon, R.\ M., Hobbs, G., \& Manchester, R.\ N. 2014, \MNRAS {\bf
    442}, 56
\bibitem[Roedig \& Sesana(2014)]{roedig:14a} Roedic, C., \& Sesana,
  A. 2014, \MNRAS {\bf 439}, 3476
\bibitem[Roedig et al.(2014)]{roedig:14} Roedig, C., Krolik, J.\ H.,
  \& Miller, M.\ C. 2014, \ApJ, {\bf 785}, 115
\bibitem[Schnittman(2013)]{schnittman:13} Schnittman, J.\ D. 2013,
  \CQG {\bf 24}, 244007
\bibitem[Schnittman et al.(2013)]{schnittman:13a} Schnittman, J.\
  D., Krolik, J.\ H., \& Noble, S.\ C. 2013, \ApJ {\bf 769}, 156
\bibitem[Schnittman \& Krolik(2014)]{schnittman:14} Schnittman,
  J.\ D., \& Krolik, J.\ H. 2014, \ApJ submitted
\bibitem[Sesana et al.(2005)]{sesana:05} Sesana, A., Haardt, F.,
  Madau, P., \& Volonteri, M. 2005, \ApJ {\bf 623}, 23
\bibitem[Sesana et al.(2007)]{sesana:07} Sesana, A., Volonteri, M., \&
  Haardt, F. 2007, \MNRAS {\bf 377}, 1711
\bibitem[Sesana et al.(2008)]{sesana:08} Sesana, A., Vecchio, A., \&
  Colacino, C.\ N. 2008, \MNRAS {\bf 390}, 192
\bibitem[Sesana et al.(2012)]{sesana:12} Sesana, A., Roedig, C.,
  Reynolds, M.\ T., \& Dotti, M. 2012, \MNRAS {\bf 420}, 860
\bibitem[Sesana(2013a)]{sesana:13a} Sesana, A. 2013a, \MNRAS {\bf 433},
  L1
\bibitem[Sesana(2013b)]{sesana:13b} Sesana, A. 2013b, \CQG {\bf 30},
  244009
\bibitem[Shi et al.(2012)]{shi:12} Shi, J.\-M., Krolik, J.\ H.,
  Lubow, S.\ H., \& Hawley, J.\ F. 2012, \ApJ {\bf 749}, 118
\bibitem[Tanaka et al.(2012)]{tanaka:12} Tanaka, T., Menou, K., \&
  Haiman, Z. 2012, \MNRAS {\bf 420}, 705
\bibitem[Tanaka \& Haiman(2013)]{tanaka:13} Tanaka, T., \&
  Haiman, Z. 2013, \CQG {\bf 30}, 224012
\bibitem[Treister et al.(2012)]{treister:12} Treister, E., Schawinski,
  K., Urry, C.\ M., \& Simmons, B.\ D. 2012, \ApJL {\bf 758}, 39 
\bibitem[Ueda et al.(2003)]{ueda:03} Ueda, Y., Akiyama, M., Ohta, K.,
  \& Miyaji, T. 2003, \ApJ {\bf 598}, 886

\end{thebibliography}
\end{document}